# Interaction of Anisotropic Dark Energy with Generalized Hybrid Expansion Law

Md Khurshid Alam, S. Surendra Singh and L. Anjana Devi

Department of Mathematics, National Institute of Technology Manipur,
Imphal-795004, India

Email: alamkhurshid96@gmail.com , ssuren.mu@gmail.com and anjnlam@gmail.com

**Abstract**: Interaction of dark energy in the anisotropic Locally Rotationally Symmetric (LRS) Bianchi type-I metric is investigated in the context of modified $f(R,T)$ theory of gravity, where $R$ is the Ricci scalar and $T$ is the trace of stress energy momentum tensor. We choose the particular form of the functional $f(R,T) = f_1(R) + f_2(T)$ then we find the exact solutions of the field equations by applying inhomogeneous equation of state, $p = \omega\rho - \Lambda(t)$ and a generalized form of hybrid expansion law. Transition of deceleration to acceleration is observed in this model. It is also observed that the Universe shows accelerated expansion at late epoch. The derived model overlaps with $\Lambda CDM$ at late time which is in agreement with present observation. Energy conditions of the derived model are also investigated. From the plot, we observe the age of Universe $(t_0 \approx 13.821 Gyr)$ for the observed $H_0 \approx 70.07 Kms^{-1}Mpc^{-1}$. The physical and geometrical behaviours of these models are also discussed.

1. **Introduction**

The examination of observational data suggests that our Universe is undergoing a phase of accelerated expansion. This result is based on the observation of high shift type SN Ia supernovae [1, 2], cosmic microwave background (CMBR) data [3] and large scales structure [4]. The observation also suggests that there had been a transition of the Universe from the earlier deceleration phase to the recent acceleration phase [5]. There are two approaches to describe this late time acceleration of Universe. One of the approaches to define the source of this acceleration is a new type of energy, which has large negative pressure, commonly known as dark energy. This dark energy has the ability to work against gravity and provide late accelerated expansion of the Universe [6]. Hence dark energy is a prime candidate to explain the accelerated expansion of the Universe. According to different measurements, dark energy contributes about 73% of the total mass energy of the Universe. While the rest 23% and 4% are occupied by dark matter, another component of the Universe which is useful in explaining the structure formation of the Universe, and baryonic matter respectively. In physical cosmology and astronomy, the simplest candidate for the dark energy is the cosmological constant $(\Lambda)$. But it has a serious problem with the fine tuning and cosmic coincidence problem, from the theoretical point of view. Due to these reasons the cosmological constant with dynamical nature is favoured over the constant cosmological constant. Therefore, numerous models of dark energy such as the quintessence phantom model, tachyon model, Chaplygin gas, Holographic dark energy, etc. have been proposed to study the mysterious nature of dark energy.

General theory of relativity is the best suited theory to explain the evolution of the Universe and $\Lambda CDM$ is the most natural model of the Universe. Modifications of general relativity are attracting more and more attention to explain late time acceleration and dark energy. Many alternative theories are developed in order to accommodate the late accelerated expansion of the Universe. Among these alternative theories, $f(R,T)$ is interesting and beneficial for study of Universe. The second approach is to modify the left-hand side of the Einstein's field equation to obtain the alternative theories of gravity. These modified gravity theories become an alternative to the conventional cosmology as it can describe the late time cosmic speed up process of expansion. Some of such theories are $f(R)$ gravity [7-9], $f(\tau)$ gravity [10,11] where $\tau$ is torsion scalar, $f(G)$ gravity [12, 13] where $G$ is Gauss-Bonnet scalar and $f(R,T)$ gravity [14] where $R$ is the Ricci scalar and $T$ is the trace of the energy momentum tensor. Among all of such theories, $f(R,T)$ gravity theory is the modification of $f(R)$ theory and became an interesting theory for the cosmologists. He also explicitly specified few classes of $f(R,T)$ modified gravity through the form of arbitrary function. Physical and geometrical properties of anisotropic cosmological model for a specific class of $f(R,T)$ are investigated [15-17].

In physical cosmology, scale factor is the key parameter to describe the different evolutionary epochs of the Universe. Power law expansion and exponential expansion of the Universe are executed with different forms of scale factor. Phase transition in the evolution of Universe from deceleration phase to acceleration phase is also observed in Universe with hybrid expansion law. In this paper we developed a generalized form of the hybrid expansion law and we also observed the transition phase of the evolution of Universe. Bianchi types IIX cosmological models becomes very interesting since these are homogeneous and anisotropic, and the process of isotropization of the Universe is investigated through the evolution of Universe. The simplicity of the field equations made Bianchi space-times useful in constructing models of spatially homogeneous and anisotropic cosmologies. Though the present Universe is homogeneous and isotropic on a larger scale, it is generally believed that the early Universe was highly anisotropic and was isotropized later with the cosmic expansion. The simplest is the Bianchi type-I Universe which is spatially homogeneous and anisotropic flat Universe. Hence many researchers have been encouraged to study the Bianchi type-I Universe as it behaves like FRW Universe at late times. Bianchi type-$VI_0$ metric is discussed with ghost dark energy and found that the model behaves like $\Lambda CDM$ at late phase of cosmic time [18]. Viscous fluid cosmological model is explored with time dependent inhomogeneous equation of state in FRW cosmological model [19]. The anisotropic cosmological model is studied in $f(R,T)$ gravity choosing the functional form of $f(R,T) = f_1(R) + f_2(T)$ and they found the cosmological parameters constrained with observational parameters [20]. Bianchi type-I metric with Hybrid Expansion Law is studied in the framework of Lyra's manifold [21]. Anisotropic cosmological models are investigated in $f(R,T)$ theory of gravity with quadratic functional form [22]. In the recent years, Bianchi Universe gains massive interest among the researchers through the observational cosmology. From the WMAP data [23], it is revealed that the standard cosmological model requires a positive and dynamic cosmological parameters that resemble the Bianchi Universe. B. Mishra et al. [24-26] also investigated cosmological models with hybrid expansion law which produce a time varying deceleration parameter that stimulates the cosmic transition.

With the motivation of the above discussions, we intend to explore a LRS Bianchi type-I cosmological model within the framework of $f(R,T)$ gravity whose functional form is $f(R,T) = \lambda R + \lambda T$, where $\lambda$ is an arbitrary constant. We solve the generalized Einstein's field equation by assuming inhomogeneous equation of state $p = \omega\rho - \Lambda(t)$ and generalized form of hybrid expansion law. This paper is organized as follows: we formulate the gravitational field equations of $f(R,T)$ theory in section 2. In Sec.3, field equations and respective solutions corresponding to $f(R,T)$ are shown. In Sec.4, we present our model with hybrid expansion law. In Sec.5, we discuss the energy conditions. Our concluding remark is given in Sec. 6.

2. **The General Formulation of $f(R,T)$ Gravity**:

By applying the Hilbert-Einstein variational action, the field equations of $f(R,T)$ theory of gravitation are derived. Harko et al. [14] use the following form of the action for $f(R,T)$ modified gravity

$$S = \frac{1}{16\pi} \int f(R,T)\sqrt{-g}\, d^4x + \int L_m \sqrt{-g}\, d^4x \qquad (1)$$

Here, $f(R,T)$ stands for arbitrary function of Ricci scalar (R) and the trace (T) of the stress energy momentum tensor. We define the stress energy momentum tensor ($T_{ij}$) as

$$T_{ij} = -\frac{2}{\sqrt{-g}} \frac{\delta(\sqrt{-g}\, L_m)}{\delta g^{ij}} \qquad (2)$$

where the matter Lagrangian density is given by $L_m$ which is governed by the metric tensor component. Therefore, stress energy momentum tensor becomes

$$T_{ij} = g_{ij} L_m - 2\frac{\partial L_m}{\partial g^{ij}} \qquad (3)$$

Taking the variation of action $S$ with respect to metric tensor $g_{ij}$, we derive the field equation of $f(R,T)$ gravity as

$$f_R(R,T)R_{ij} - \frac{1}{2} f(R,T) g_{ij} + (g_{ij}\Box - \nabla_i \nabla_j) f_R(R,T) = 8\pi T_{ij} - f_T(R,T)T_{ij} - f_R(R,T)\Theta_{ij} \qquad (4)$$

where $\Theta_{ij} = g^{\alpha\beta} \dfrac{\delta T_{\alpha\beta}}{\delta g^{ij}}$. Using the value of the matter Lagrangian $L_m$, and using the value of $T_{ij}$, we have

$$\Theta_{ij} = -2T_{ij} + g_{ij} L_m - 2g^{\alpha\beta} \frac{\partial^2 L_m}{\partial g^{ij} \partial g^{lm}} \qquad (5)$$

We consider the perfect fluid model governed by energy density $\rho$ and pressure $p$ of the fluid with four-velocity $u^i = (0,0,0,1)$ satisfying $u^i u_j = 1$ and $u^i \nabla_j u_j = 0$. We take the energy momentum tensor of perfect fluid in the form

$$T_{ij} = (\rho + p)u_i u_j - p g_{ij} \tag{6}$$

We take the matter Lagrangian $L_m = -p$. Then equation (5) becomes

$$\Theta ij = -2Tij - pgij \tag{7}$$

The field equations of $f(R,T)$ theory of gravity also related to the physical behaviours of the matter field through the tensor $\Theta ij$. Harko et al. [14] introduced the three classes which are given as

$$f(R,T) = \begin{cases} R + 2f(T) \\ f_1(R) + f_2(T) \\ f_1(R) + f_2(R) f_3(T) \end{cases}$$

3. **Metric and Field Equations:**

In this paper, we choose the case $f(R,T) = f_1(R) + f_2(T)$ for LRS Bianchi type-I metric to develop our model. The LRS Bianchi type-I metric of this model is given as

$$ds^2 = dt^2 - A^2(t)(dx^2 + dy^2) - B^2(t)dz^2 \tag{8}$$

Here $A$ and $B$ are metric potentials which are functions of time. This metric has symmetric plane in $xy$-plane and symmetry axis are along $z$-axis. The volume of the Universe is given by $V = A^2 B = a^3$, where $a$ is the scale factor. And the Hubble parameter is given by

$$H = \frac{1}{3}(2H_x + H_z) \tag{9}$$

where $H_x = H_y = \frac{\dot{A}}{A}$ and $H_z = \frac{\dot{B}}{B}$.

Then we take $f(R,T) = \lambda R + \lambda T$, where $\lambda$ is an arbitrary constant. Setting $(g_{ij}\Box - \nabla_i \nabla_j) = 0$, equation (4) reduces to

$$R_{ij} - \frac{1}{2}Rg_{ij} = \left(p + \frac{1}{2}T\right)g_{ij} + \left(\frac{8\pi + \lambda}{\lambda}\right)T_{ij} \tag{10}$$

Using the energy momentum tensor of the matter Lagrangian for perfect fluid obtained in equation (6) and field equations obtained in equation (10), the expressions of gravitational field equations for the metric in equation (8) are calculated by

$$\left(\frac{\dot{A}}{A}\right)^2 + 2\frac{\dot{A}}{A}\frac{\dot{B}}{B} = \left(p + \frac{1}{2}T\right) + \alpha\rho \tag{11}$$

$$\frac{\ddot{A}}{A} + \frac{\ddot{B}}{B} + \frac{\dot{A}}{A}\frac{\dot{B}}{B} = \left(p + \frac{1}{2}T\right) + \alpha p \tag{12}$$

$$2\frac{\ddot{A}}{A} + \left(\frac{\dot{A}}{A}\right)^2 = \left(p + \frac{1}{2}T\right) - \alpha p \tag{13}$$

where an overhead dot and the double overhead dots hereafter denote the first and second differentiation with respect to cosmic time '$t$' respectively and $\alpha = \frac{8\pi\lambda + \lambda}{\lambda}$. The trace of the stress energy momentum tensor $T$ in our derived model is given by $T = \rho - 3p$. We consider a Universe filled with anisotropic dark energy obeying the inhomogeneous equation of state in the form $p = \omega\rho - \Lambda(t)$, where $-1 \leq \omega \leq 1$ and $\Lambda(t)$ is time dependent cosmological constant. This equation of state, with $w(t)$ a function of time, was examined in [27]. The above field equations (11)-(13) can be written in terms of $H_x$ and $H_z$ as

$$H_x^2 + 2H_xH_z = \frac{1}{2}\Lambda + \frac{1}{2}\rho(1 - \omega + 2\alpha) \tag{14}$$

$$\dot{H}_x + \dot{H}_z + H_x^2 + H_z^2 + H_xH_z = \left(\frac{1}{2} + \alpha\right)\Lambda + \frac{1}{2}\rho(1 - \omega - 2\omega\alpha) \tag{15}$$

$$2\dot{H}_x + 3H_x^2 = \left(\frac{1}{2} + \alpha\right)\Lambda + \frac{1}{2}\rho(1 - \omega - 2\omega\alpha) \tag{16}$$

On solving equations (12)-(13), we get

$$A = c_2^{\frac{1}{3}} a \exp\left(\frac{c_1}{3}\int \frac{dt}{a^3}\right) \tag{17}$$

$$B = c_2^{-\frac{2}{3}} a \exp\left(\frac{-2c_1}{3}\int \frac{dt}{a^3}\right) \tag{18}$$

where $c_1$ and $c_2$ are constants of integration. The cosmological term, energy density and pressure are obtained from eqs. (14)-(16)

$$\Lambda = \frac{1}{\alpha(\alpha+1)}[(\omega\alpha - \omega + 3\alpha + 1)H_x^2 + (\omega + 2\alpha - 1)\dot{H}_x + (2\omega\alpha + \omega - 1)H_xH_z] \tag{19}$$

$$\rho = \frac{1}{\alpha(\alpha+1)}[(\alpha - 1)H_x^2 - \dot{H}_x + (2\alpha + 1)H_xH_z] \tag{20}$$

$$p = -\frac{1}{\alpha(\alpha+1)}[(3\alpha + 1)H_x^2 + (2\omega + 2\alpha - 1)\dot{H}_x - H_xH_z] \tag{21}$$

4. **Model with Generalized Hybrid Expansion Law:**

Many authors [24-26] also investigated cosmological models with hybrid expansion which stimulates the cosmic transition from deceleration to acceleration epoch. In order to obtain generalized hybrid Universe, we consider a generalized hybrid scale factor of the Universe in the form

$$a(t) = b_0 t^m e^{t^{1-m}} \tag{22}$$

where $b_0 \geq 0$ and $m \geq 0$ are constants. This form of scale factor is called generalized hybrid expansion law (GHEL). In particular cases, one can obviously obtains power law and exponential expansions of the Universe with $m = 1$ and $m = 0$ respectively. Thus power law and exponential law cosmology are special cases of the hybrid expansion law cosmology which is in the valid interval $0 < m < 1$. This hybrid expansion law also provides the transition phase of deceleration phase to accelerating phase of Universe as in [24-26]. Volume of the Universe (V) is given by $V = a^3(t)$. This gives the scale factor as

$$a(t) = b_0^{\frac{1}{3}} t^{\frac{m}{3}} e^{\frac{t^{1-m}}{3}} \qquad (23)$$

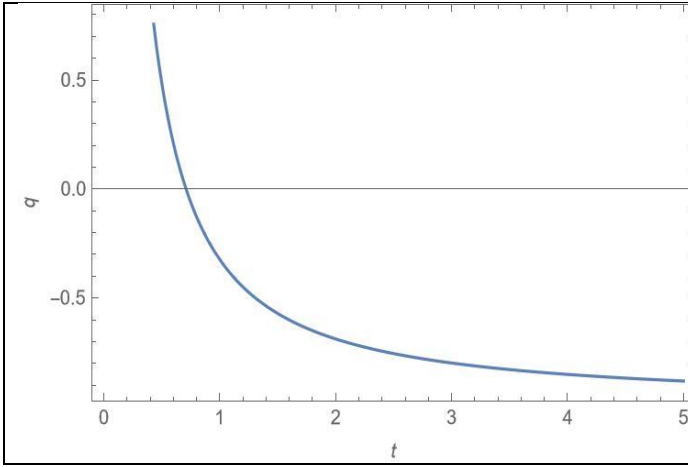
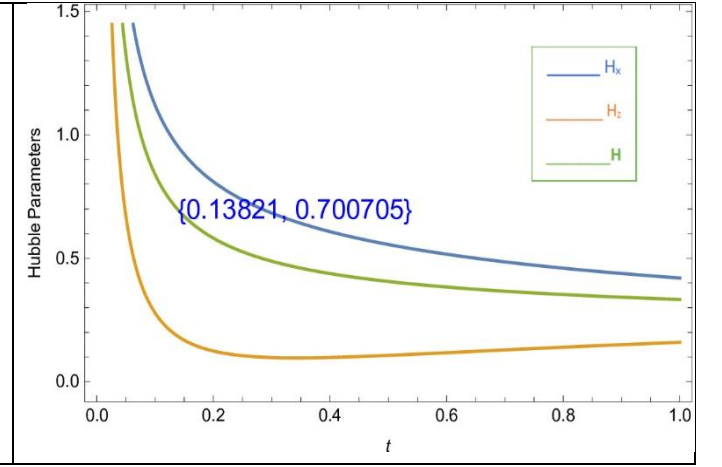

Fig.1    Fig.2

Fig. 1: Time variation of deceleration parameter (q). Fig. 2: Time variation of Hubble parameter and Hubble directional parameter.

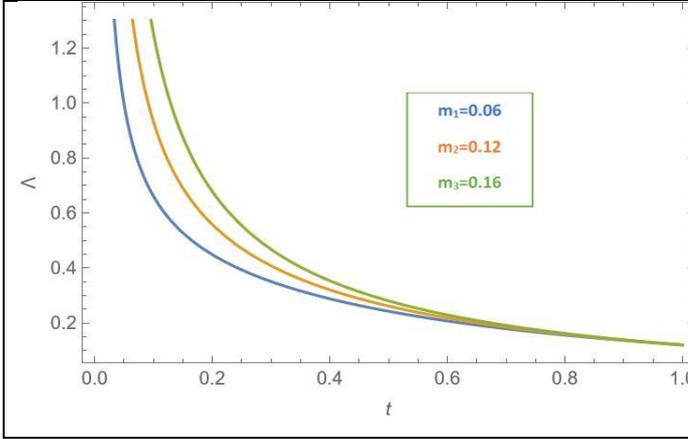
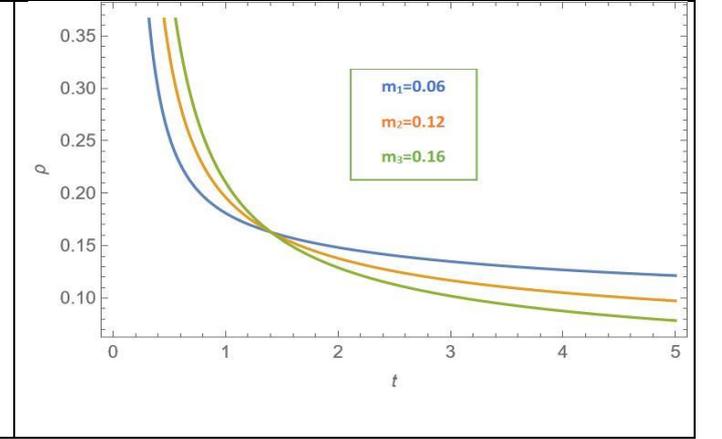

Fig.3    Fig.4

Fig. 3: Time variation of Cosmological constant $(\Lambda)$. Fig. 4: Time variation of density $(\rho)$.

For this scale factor, the deceleration parameter is found to be

$$q = -1 + \frac{3m\{t^{-2} + (1-m)t^{-(m+1)}\}}{[mt^{-1} + (1-m)t^{-m}]^2} \qquad (24)$$

Here $q$ approaches to $-1$ which shows that the present Universe approaches to the epoch of accelerating expansion as $t \to \infty$. Using equation (23) in (17) and (18) we obtain

$$A = \left( c_2 b_0 e^{\frac{c_1 c_3}{b_0}} \right)^{\frac{1}{3}} t^{\frac{m}{3}} e^{\frac{t^{1-m}}{3} - \frac{c_1 e^{-t^{1-m}}}{3b_0(1-m)}} \qquad (25)$$

$$B = \left( c_2^{-2} b_0 e^{\frac{-2c_1 c_3}{b_0}} \right)^{\frac{1}{3}} t^{\frac{m}{3}} e^{\frac{t^{1-m}}{3} + \frac{2c_1 e^{-t^{1-m}}}{3b_0(1-m)}} \qquad (26)$$

This shows that the scale factors expand along $x$, $y$ and $z$ axes with different rates of expansion. Also the expression of directional Hubble parameters, Hubble parameter, anisotropy parameter, expansion scalar and the shear scalar becomes

$$H_x = \frac{m}{3}t^{-1} + \frac{1-m}{3}t^{-m} + \frac{c_1}{3b_0}e^{-t^{1-m}}t^{-m} \tag{27}$$

$$H_z = \frac{m}{3}t^{-1} + \frac{1-m}{3}t^{-m} - \frac{2c_1}{3b_0}e^{-t^{1-m}}t^{-m} \tag{28}$$

$$H = \frac{m}{3}t^{-1} + \frac{1-m}{3}t^{-m} \tag{29}$$

$$\Delta = \frac{1}{3}\sum_{i=1}^{3}\left(\frac{H_i - H}{H}\right)^2 = \frac{2c_1^2}{b_0^2}\left[\frac{e^{-t^{1-m}}t^{-m}}{mt^{-1}+(1-m)t^{-m}}\right]^2 \tag{30}$$

$$\theta = 3H = mt^{-1} + (1-m)t^{-m} \tag{31}$$

$$\sigma^2 = \frac{1}{2}\left(\sum_{i=1}^{3}H_i^2 - 3H^2\right) = \frac{c_1^2}{3b_0^2}e^{-2t^{1-m}}t^{-2m} \tag{32}$$

Using the values of $H_x$ and $H_z$ in equation (19)-(21), we obtain the cosmological term, energy density and pressure as follow:

$$\Lambda = \frac{1}{3\alpha(\alpha+1)}[\{\alpha(\omega+1)m^2 - (\omega+2\alpha-1)m\}t^{-2} + \{\alpha(\omega+1)(1-m)^2 - (\alpha+1)(\omega+1)\frac{c_1^2}{b_0^2}e^{-2t^{1-m}}$$

$$-2(\omega-1)(1-m)\frac{c_1}{b_0}e^{-t^{1-m}}\}t^{-2m} + \{(2\omega\alpha-w+1)m(1-m) - 2(\omega-1)\frac{mc_1}{b_0}e^{-t^{1-m}}\}t^{-(m+1)}] \tag{33}$$

$$\rho = \frac{1}{3\alpha(\alpha+1)}[m(m\alpha+1)t^{-2} + \{\alpha(1-m)^2 - (\alpha+1)\frac{c_1^2}{b_0^2}e^{-2t^{1-m}}\}t^{-2m} + (2\alpha+1)m(1-m)t^{-(m+1)}] \tag{34}$$

$$p = -\frac{1}{3\alpha(\alpha+1)}[\{\alpha m^2 - (2\omega+2\alpha-1)m\}t^{-2} + \{\alpha(1-m)^2 + (\alpha+1)(\omega+1)\frac{c_1^2}{b_0^2}e^{-2t^{1-m}}$$

$$+2(1-\omega)(1-m)\frac{c_1}{b_0}e^{-t^{1-m}}\}t^{-2m} + \{(1-2\omega)m(1-m) + 2(1-\omega)\frac{mc_1}{b_0}e^{-t^{1-m}}\}t^{-(m+1)}] \tag{35}$$

Here we plot the dark energy model ($\omega = -1$) for $\alpha = 1, c_1 = 0.85, b_0 = 1.2$ and $m = 0.06, 0.12, 0.16$. It is observed that the deceleration parameter $q$ is positive at early stage of the Universe and negative for late time Universe for $0 < m < 1$ which indicates that the Universe exhibits transition phase from deceleration to acceleration. The nature of the deceleration parameter is shown in Figure 1. From Figure 2, it is evident that the directional Hubble parameters are exceptionally large at the beginning of the Universe and decrease monotonically with its age. Hubble constant ($H_0$) is measured to be about $70 - 76 Kms^{-1}Mpc^{-1}$ by a variety of techniques. The best current results using Cepheids and the Hubble Space Telescope come from the SHOES team which measure a value of about $73.5 Kms^{-1}Mpc^{-1}$. Recent measurements based on red giant stars give a value of $70 - 72 Kms^{-1}Mpc^{-1}$. Recently, Plank 2018 results. VI shows the present value of Hubble constant ($H_0$) as $(67.4 \pm 0.5) Kms^{-1}Mpc^{-1}$ and present age of Universe to be $13.802 \pm 0.024 Gyr$. From this plot of Hubble constant vs Time in Figure 2 for the valid values $m = 0.12$, we obtain the age of Universe $(t_0 \approx 13.821 Gyr)$ for the observed $H_0 \approx 70.07 Kms^{-1}Mpc^{-1}$ which is in agreement with the observational values of [28]. By using the value of $t_0$ in equation (24) for $m = 0.134$, we calculated the values of present deceleration parameter $(q_0)$ as $q_0 \approx -1$ which is in agreement with present observation. In Figure 5, we see that the anisotropic parameter approaches zero as $t$ tends to infinity. It means that our Universe approaches isotropy at late epoch of its evolution. These scenarios provide information that our Universe is highly anisotropic in the past and becomes isotropic later. The expansion scalar shows that the Universe initially evolves with an infinite expansion and decreases monotonically at late times. It is evident that the cosmological constant term $\Lambda$ is positive and becomes very small as shown in the Figure 3. Also the energy density is positive and is a decreasing function of time from the Figure 4.

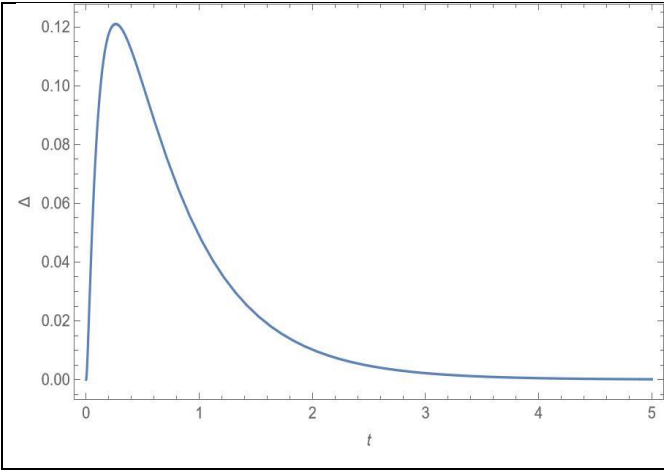 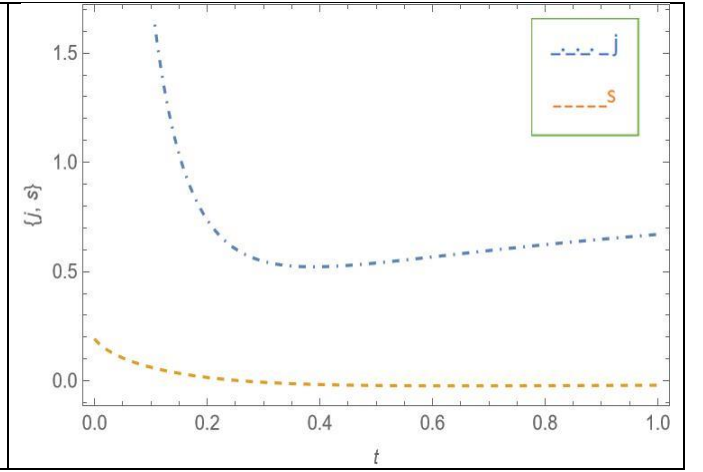

Fig. 5                                                    Fig.6
Fig. 5: Time variation of anisotropic parameter $(\Delta)$. Fig. 6: Time variation of $\{j,s\}$.

5. **Energy Conditions and Some Observational Parameters:**

For observational investigation, we consider $a(t) = 1/1 + z$, where $z$ is the redshift. This relation gives the expression between time and redshift as

$$t = \left(\frac{m}{1-m}\right)^{\frac{1}{1-m}} W\left[\frac{1-m}{m}\left(\frac{1}{b_0(1+z)^3}\right)^{\frac{1-m}{m}}\right]^{\frac{1}{1-m}} \qquad (36)$$

where $W$ denotes the Lambert $W$ function. Using (23), we can be expressed the parameters of the derived model in terms of the redshift. Such a relation is useful for testing the model with observational data. The essence of dark energy models can be found out through the state finder diagnostic pair $\{j,s\}$ which give us an idea about the geometrical nature of the model. The pair $\{j,s\}$ are defined as [29,30]

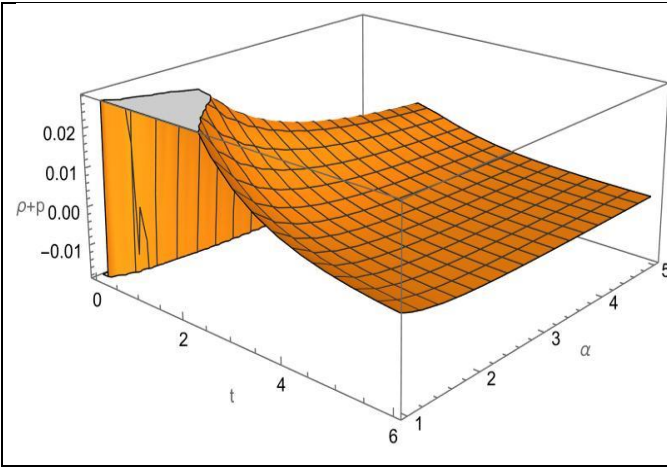 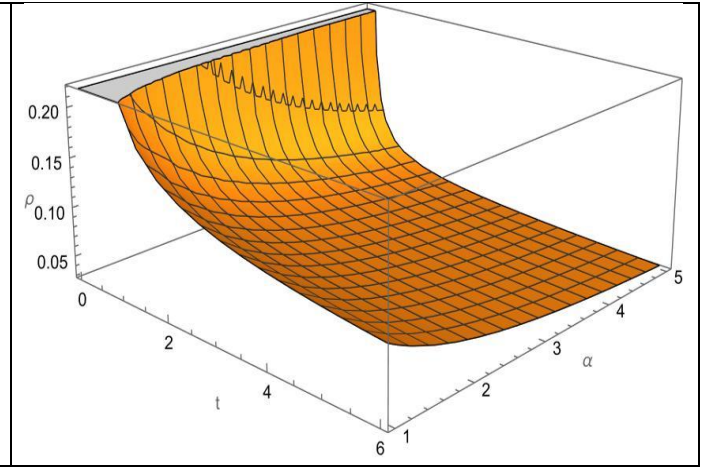

Fig.7　　　　　　　　　　　　　　　　　　　　Fig.8

Fig. 7: Plot of $\rho + p \geq 0$ vs $\alpha$ and time (t). Fig. 8: Plot of $\rho \geq 0$ vs $\alpha$ and time (t).

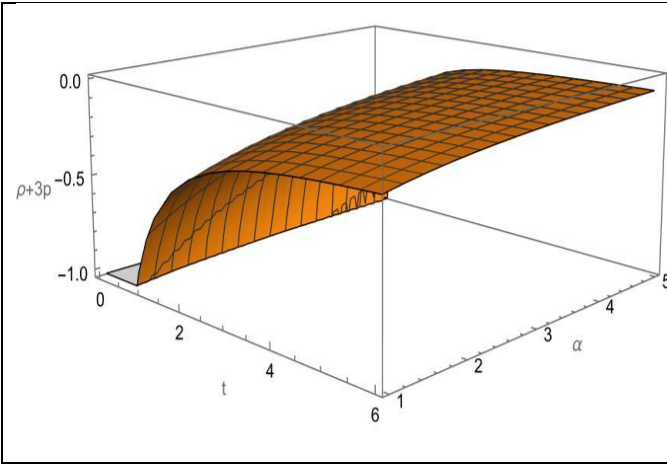 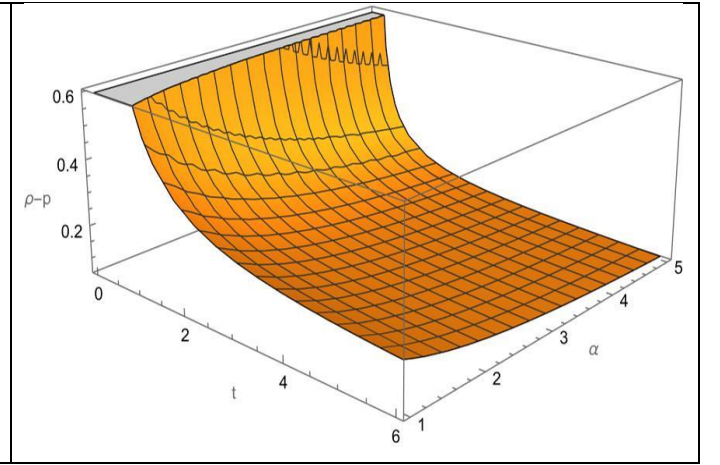

Fig.9　　　　　　　　　　　　　　　　　　　　Fig.10

Fig. 9: Violation of $\rho + 3p \geq 0$ vs $\alpha$ and time (t). Fig. 10 Plot of $\rho - p \geq 0$ vs $\alpha$ and time (t).

$$j = \frac{\dddot{a}}{aH^3} = 1 + \frac{\left(\frac{-m}{9} + \frac{2m}{9}\right)t^{-3} - 6m^2(1-m)t^{-(m+2)} - 3m(1-m)^2 t^{-(2m+1)}}{\{mt^{-1} + (1-m)t^{-m}\}^3} \quad (37)$$

$$s = \frac{j-1}{3\left(q - \frac{3}{2}\right)} = \frac{2\left(\frac{-m}{9} + \frac{2m}{9}\right)t^{-3} - 6m^2(1-m)t^{-(m+2)} - 3m(1-m)^2 t^{-(2m+1)}}{3\{mt^{-1} + (1-m)t^{-m}\}\{-5mt^{-1} - 5(1-m)t^{-m} + 6mt^{-2} + 6m(1-m)t^{-(m+1)}\}} \quad (38)$$

In the above definition of $s$, there is $\frac{3}{2}$ in the place of $\frac{1}{2}$ in the original definition $s = \frac{j-1}{3\left(q - \frac{1}{2}\right)}$ [28] to avoid divergence of $s$ when $q = \frac{1}{2}$. Figure 6 shows that the model behaves like $\Lambda CDM$ model for $n \to \infty$ as $\{j, s\}$ becomes $(1, 0)$ for $0 < m < 1$. The energy conditions in general relativity are null energy condition (NEC), weak energy condition (WEC), strong energy condition (SEC) and dominant energy condition (DEC) which are expressed respectively as

$$NEC \Leftrightarrow \rho + p \geq 0 \quad (39)$$
$$WEC \Leftrightarrow NEC \text{ and } \rho \geq 0 \quad (40)$$
$$SEC \Leftrightarrow \rho + 3p \geq 0 \quad (41)$$
$$DEC \Leftrightarrow \rho - p \geq 0 \quad (42)$$

From Figures (7)-(10), we observe that $\rho + p > 0$, $\rho > 0$, $\rho + 3p < 0$ and $\rho - p > 0$. In this model, the null energy condition (NEC), weak energy condition (WEC) and dominant energy condition (DEC) are satisfied but the strong energy condition (SEC) is violated. The same is suggested by various researchers [31, 32].

6. **Conclusion:**

In this paper, we have established the solution of Einstein's field equations for LRS Bianchi-I spacetime using inhomogeneous equation of state $p = \omega\rho - \Lambda(t)$. In this model, $V \to \infty$ as $t \to \infty$ which represents the accelerated expansion of the Universe. The physical behaviours of the dynamic cosmological parameters depend on the value of $m$ and we discuss the model with Hybrid expansion for $0 < m < 1$. The expansion scalar shows that the expansion rate is infinite at the beginning and decreases monotonically at late times. As the deceleration parameter changes from positive to negative value with time, model exhibits a transition phase from deceleration to acceleration, which is an important feature for the evolution of Universe. It is fascinating to find that the model exhibits an initial singularity with high anisotropy. The anisotropic parameter approaches zero as $t$ tends to infinity which shows that our Universe approaches isotropy at late times. The cosmological constant term and the energy density decrease monotonically with time which is in agreement with observations. Thus our model becomes spatially homogeneous, isotropic and flat at late times. Furthermore it is found that the values of state finder pair becomes $(j = 1, s = 0)$ at late epoch. This shows that our derived model approaches to the $\Lambda CDM$ model at late epoch. From the Figure 2, we obtain age of Universe $(t_0 \approx) 13.821 Gyr$ for the observed $H_0 \approx 70.07 Kms^1 Mpc^{-1}$ which is in agreement with the observations. By using the value of $t_0$ in equation (24) for $m = 0.12$, we calculated $(q_0)$ as $q_0 \approx -1$ which is in agreement with present observation. Also in this model, the strong energy condition (SEC) is violated while null energy conditions (NEC), weak energy conditions (WEC) and dominant energy condition (DEC) are satisfied which are agreeable with the present time. Although the models obtained here are simple, it may be useful in the investigation of the evolution of the Universe.

**Acknowledgement**: We would like to thank the National Institute of Technology Manipur for financial support.